\documentstyle[preprint,prb,aps]{revtex}

\begin{document}
\draft
\title{The ubiquitous XP commutator}
\author{Alberto C. de la Torre}
\address{Departamento de F\'{\i}sica,
 Universidad Nacional de Mar del Plata\\
 Funes 3350, 7600 Mar del Plata, Argentina\\
dltorre@mdp.edu.ar}
\maketitle
\begin{abstract}
Position and momentum of a particle can take any value in a
continuous spectrum; these values are independent but their
indeterminacies are correlated; momentum and position are
mutually the generators of the transformations in each other. It
is shown in a concise way, how all these features arise solely
from their commutation relation $[X,P]=i\hbar$. The article is
complete and self contained, adequate for didactic use.
\end{abstract}

\section{INTRODUCTION}
According to quantum mechanics there are four outstanding
features concerning the observables of position $X$ and momentum
$P$ of a particle. They are:
\begin{itemize}
    \item The spectrum of position and momentum operators is
    \emph{continuous} and can take any real value. That is, physically,
    the result of measuring position or momentum of a
    particle can assume any real value. This is of course not
    surprising since it is also true in classical mechanics,
    however it is not a trivial statement because in quantum
    mechanics there are many observables with non continuous, that is,
    discrete spectra.
    \item Position and momentum in quantum mechanics, as well as
    in classical mechanics, are independent. Physically, this means
    that at a given position of a particle, any value of momentum
    is possible without any preference, and viceversa, a given
    momentum can be realized at any position. In classical
    mechanics we take then two independent variables to describe
    position and momentum and all other relevant observables are
    functions of these two independent variables. In quantum mechanics this
    independence must be formalized in a more subtle way: to have
    a particle at a given position $x$ means that the system is
    described by a state $\varphi_{x}$, an eigenvector of
    the position operator $X$ corresponding to the eigenvalue
    $x$, that is, $X\varphi_{x}=x\varphi_{x}$. If, in this state,
    any value $p$ of momentum is equally probable, then the
    probability $|\langle\varphi_{x},\phi_{p}\rangle|^{2}$ must
    be independent of $p$, where $\phi_{p}$ is the eigenvector
    of the momentum operator associated to the eigenvalue $p$,
    that is, $P\phi_{p}=p\phi_{p}$. Similarly, if the particle has
    some value of momentum $p$, its state is $\phi_{p}$ and in
    order to have any position $x$ equally probable we must
    have $|\langle\phi_{p},\varphi_{x}\rangle|^{2}$ independent
    of $x$. Now the two sets $\{\varphi_{x},\forall x\}$
    and $\{\phi_{p},\forall p\}$ are two different bases of the
    Hilbert space and they are called \emph{mutually unbiased} if the
    number $|\langle\varphi_{x},\phi_{p}\rangle|$ is a constant
    independent of $x$ and $p$. Therefore the independence of
    position and momentum in quantum mechanics is formalized by
    the requirement that their corresponding bases should be
    unbiased.
    \item The values of position and momentum are independent but
     their indeterminacies in any state $\psi$, given by
     $\Delta_{x}^{2}=\langle\psi,(X-\langle X\rangle)^{2}\psi\rangle$
     and $\Delta_{p}^{2}=\langle\psi,(P-\langle P\rangle)^{2}\psi\rangle$,
     where $\langle X\rangle=\langle\psi,X\psi\rangle$ and
     $\langle P\rangle=\langle\psi,P\psi\rangle$ are the expectation values
      of position and momentum,
     are \emph{correlated} such that $\Delta_{x}\Delta_{p}\geq\hbar/2$.
     This feature is unique to quantum mechanics since there are
     no indeterminacies in classical mechanics. The correlation
     manifest in the uncertainty principle is unavoidable and
     appears in \emph{all states}; however for \emph{some states}, the
     lower bound in the inequality can take a larger value as we
     will see later.
    \item Momentum and position observables are \emph{generators}
    of translations in space and of increases in momentum. This
    relation, also present in classical mechanics, is formalized
    in quantum mechanics by the definition of a unitary operator
    $U_{a}=\exp(-\frac{i}{\hbar}aP)$ that performs the
    transformations of states and operators in the Hilbert space
    corresponding to the translations in physical space
    $X\rightarrow X+a$. So for any function of the position operator
   we have $F(X+a)=U^{\dag}_{a}\ F(X)\ U_{a}$. Similarly, the
   operator $V_{g}=\exp(+\frac{i}{\hbar}gX)$ describes in the
   Hilbert space the boosts\cite{boost} $P\rightarrow P+g$, and
   $G(P+g)=V^{\dag}_{g}\ G(P)\ V_{g}$.
\end{itemize}

A beautiful feature of the formalism of quantum mechanics is that
al properties mentioned above can be shown to be consequences of
the commutation relation $[X,P]=i\hbar$ solely. The commutator is
indeed ubiquitous and contains encoded all relevant properties of
position, momentum and their relations.  We will give a complete
proof of the above statements, although some parts are very well
known and appear in most textbooks. We do this because in some
cases the proof presented here have some advantages compared to
the standard ones (for instance of the uncertainty relations) and
some other parts of the proof are absent from most textbooks,
even from advanced ones.
\section{INDETERMINACY RELATIONS}
Let us start presenting a proof\cite{dlt1} of the indeterminacy
relations in the optimized version given by
Schr{\"o}dinger\cite{schro} three years after the appearance of
Heisenberg's uncertainty relations\cite{heis}. For this proof, it
is convenient to work with the ``centered'' operators defined as
$\widetilde{X}=X-\langle X\rangle$ and $\widetilde{P}=P-\langle
P\rangle$ (of course, the expectation values are multiplying an
identity operator not shown) and therefore we have, for any state
$\psi$, $ \Delta_{x}^{2}=\langle\psi,\widetilde{X}^{2}\psi\rangle
= \langle\widetilde{X}\psi,\widetilde{X}\psi\rangle =
\|\widetilde{X}\psi\|^{2}$ and similarly $
\Delta_{p}^{2}=\langle\psi,\widetilde{P}^{2}\psi\rangle =
\langle\widetilde{P}\psi,\widetilde{P}\psi\rangle =
\|\widetilde{P}\psi\|^{2}$. Their product is then
\begin{equation}\label{indet1}
\Delta_{x}^{2}\Delta_{p}^{2}
=\left(\|\widetilde{X}\psi\|\|\widetilde{P}\psi\|\right)^{2} \geq
\left|\langle\widetilde{X}\psi,\widetilde{P}\psi\rangle\right|^{2}
=\left|\langle\psi,
\widetilde{X}\widetilde{P}\psi\rangle\right|^{2}\ .
\end{equation}
The inequality sign comes from Cauchy-Schwarz inequality for the
internal product $|\langle f,g\rangle|\leq\|f\|\|g\|$. Now we can
write the product $\widetilde{X}\widetilde{P}
=\frac{1}{2}\{\widetilde{X}\widetilde{P}\} + i\frac{1}{2i}
[\widetilde{X}\widetilde{P}] $. In this way, the product is
decomposed in the hermitian part plus the antihermitian part
(similar to the decomposition of a complex number in real and
imaginary part), and therefore we have
\begin{equation}\label{indet2}
\Delta_{x}^{2}\Delta_{p}^{2} \geq  \left(\langle\psi,
\frac{1}{2}\{\widetilde{X},\widetilde{P}\}\psi\rangle\right)^{2}
+\left(\langle\psi,
\frac{1}{2i}[\widetilde{X},\widetilde{P}]\psi\rangle\right)^{2}\
.
\end{equation}
Now we replace the commutator and anticommutator
\begin{eqnarray}
  [\widetilde{X},\widetilde{P}] &=& [X,P] = i\hbar \\
 \{\widetilde{X},\widetilde{P}\} &=& \{X,P\} -2 X\langle P\rangle-2
 \langle  X\rangle P + 2 \langle  X\rangle \langle P\rangle\ ,
\end{eqnarray}
and we finally get the indeterminacy relation
\begin{equation}\label{indet3}
\Delta_{x}^{2}\Delta_{p}^{2} \geq \left(
\frac{\hbar}{2}\right)^{2} +\left(\frac{1}{2}\langle\psi,
(XP+PX)\psi\rangle-\langle X\rangle \langle P\rangle\right)^{2}\
.
\end{equation}
The second parenthesis vanishes for some states (for instance
gaussian packets) and it is ignored in  many texts because it is
nonnegative. However, for other states, it gives an inequality
more restrictive than the original Heisenberg's relation
$\Delta_{x}\Delta_{p}\geq\hbar/2$.
\section{GENERATORS}
We will now prove that the commutation relation implies that the
momentum $P$ is the generator of translations $X\rightarrow X+a$.
Precisely, we will show that if $[X,P] = i\hbar $ then for any
function (that can be expanded as a power series) $F(X)$ it is
\begin{equation}\label{gen1}
   F(X+a)=U^{\dag}_{a}\ F(X)\ U_{a} \ ,
\end{equation}
where the unitary operator $U_{a}$ is given by
\begin{equation}\label{gen2}
   U_{a}=\exp(-\frac{i}{\hbar}aP)\ ,
\end{equation}
and if $\varphi_{x}$ is an eigenvector of the position operator
$X$ corresponding to the eigenvalue $x$, then $U_{a}\varphi_{x}$
is an eigenvector corresponding to the eigenvalue $x+a$.

In order to prove Eq.(\ref{gen1}) we will consider the Taylor
series expansion of both sides of the equation. The left hand
side will be expanded as a function of $X$ and the expansion will
involve $n$th order derivatives of $F(X)$, whereas the right hand
side will be expanded as a function of the parameter $a$ and it
will involve derivatives with respect to $a$ that, due  to the
rule for the derivative of a product, will be given by an $n$th
order commutator. The proof is obtained by the identification of
the derivatives of $F(X)$ with its commutator with $P$. Let us
begin by this last step: If $[X,P] = i\hbar$, then for any
function $F(X)$ (that can be expanded as a power series) it is
\begin{equation}\label{gen3}
   \frac{d^{n}F(X)}{dX^{n}}= \frac{-i}{\hbar}[\frac{-i}{\hbar}[
   \cdots\frac{-i}{\hbar}[F(X),P],P],\cdots,P]
   =\left(\frac{-i}{\hbar}[\right)^{n}F(X)\left(,\frac{}{}P]\right)^{n}\ ,
\end{equation}
where the ``$n$ exponent'' means an $n$ fold repetition of the
enclosed symbols. In order to prove this we first notice that
\begin{equation}\label{gen4}
  [X^{n},P]=i\hbar nX^{n-1} = \frac{d}{dX}i\hbar X^{n} \ ,
\end{equation}
that can be proven by mathematical induction: for $n=1$ it is
valid, assuming that it is valid for $n$, we prove that it is
valid for $n+1$. Indeed
$[X^{n+1},P]=X[X^{n},P]+[X,P]X^{n}=Xi\hbar nX^{n-1} + i\hbar
X^{n} = i\hbar (n+1)X^{n}$. Now we take an arbitrary set of
coefficients $\{C_{n}\}$ to multiply  Eq.(\ref{gen4}) and sum
over $n$. So we get for any function $F(X)=\sum_{n}C_{n}X^{n}$
\begin{equation}\label{gen5}
  [F(X),P]= i\hbar\frac{dF(X)}{dX}  \ .
\end{equation}
If we iterate this result, for the $n$th order derivative we
obtain Eq.(\ref{gen3}).

Let us consider now he right hand side of Eq.(\ref{gen1}) as a
function of $a$ and its derivatives.
\begin{eqnarray}
  f(a) &=& \exp(\frac{i}{\hbar}aP)\ F(X)\ \exp(\frac{-i}{\hbar}aP)
\nonumber \\
  \frac{df(a)}{da} &=& \frac{i}{\hbar}Pf(a)+f(a)\frac{-i}{\hbar}P
  =\frac{-i}{\hbar}[f(a),P]\nonumber \\
  \frac{d^{2}f(a)}{da^{2}} &=&\frac{-i}{\hbar}[ \frac{-i}{\hbar}[
   f(a),P],P]\nonumber \\ \cdots &\cdots& \cdots \nonumber \\
  \frac{d^{n}f(a)}{da^{n}} &=& \left(\frac{-i}{\hbar}[\right)^{n}f(a)
  \left(,\frac{}{}P]\right)^{n}\ .
\end{eqnarray}
With these derivatives, evaluated at $a=0$, we obtain the Taylor
expansion
\begin{equation}\label{gen6}
  f(a)=\left.\sum_{n=0}^{\infty}\frac{a^{n}}{n!}\frac{d^{n}f(a)}{da^{n}}
  \right|_{a=0} =\sum_{n=0}^{\infty}\frac{a^{n}}{n!}
  \left(\frac{-i}{\hbar}[\right)^{n}F(X)
  \left(,\frac{}{}P]\right)^{n}\ .
\end{equation}
But now, from Eq.(\ref{gen3}) the $n$th order commutator is
replaced by the $n$th order derivative of $F(X)$ and we obtain
precisely the Taylor expansion of $F(X+a)$:
\begin{equation}\label{gen7}
  f(a)=\sum_{n=0}^{\infty}\frac{a^{n}}{n!}\frac{d^{n}F(X)}{dX^{n}}
   =F(X+a)\ .
\end{equation}

This completes the proof of Eq.(\ref{gen1}). As a particular case
of this result, when the function $F(X)$ is the identity, we have
$X+a=U^{\dag}_{a}\ X\ U_{a} $ and performing a left-product with
$U_{a}$ we obtain the commutator
\begin{equation}\label{gen8}
  [X,U_{a}]= aU_{a} \ .
\end{equation}
With this commutator we can easily prove that $U_{a}$ is a shift
operator for the eigenvectors of the position operator.
\begin{equation}\label{gen9}
 X U_{a}\varphi_{x}= U_{a}X\varphi_{x}+aU_{a}\varphi_{x}
 =(x+a)U_{a}\varphi_{x}\ ,
\end{equation}
therefore (assuming nondegeneracy)
\begin{equation}\label{gen10}
 U_{a}\varphi_{x}= \varphi_{x+a}\ .
\end{equation}

We have proven that $P$ is the generator of translations
$X\rightarrow X+a$ and exactly in the same way we can prove that
$X$ is the generator of boosts $P\rightarrow P+g$ with the
difference in the sign of the exponent in the unitary operator
$V_{g}=\exp(+\frac{i}{\hbar}gX)$. Then we have
$G(P+g)=V^{\dag}_{g}\ G(P)\ V_{g}$, and if $\phi_{p}$ is an
eigenvector of the position operator $P$ corresponding to the
eigenvalue $p$, then $V_{g}\phi_{p}=\phi_{p+g}$ is an eigenvector
corresponding to the eigenvalue $p+g$.

Another interesting consequence or Eq.(\ref{gen1}) follows from
the special case where we take
$F(X)=V_{g}=\exp(+\frac{i}{\hbar}gX)$. With this we get:
$\exp(\frac{i}{\hbar}ga)V_{g}=U^{\dag}_{a}V_{g}U_{a}$, and from
this we obtain the commutation relation for the unitary
transformations corresponding to translations and boosts
\begin{equation}\label{comm}
  [V_{g},U_{a}]=\left(\exp(\frac{i}{\hbar}ga)-1\right)
U_{a}V_{g}\ .
\end{equation}
Although $X$ and $P$ do  not commute, boosts and translations
commute when the parameters are such that $ag=2\pi\hbar m$ for
any integer $m$. This interesting relation was derived for solid
state physics\cite{zak} where $a$ is a lattice constant and $g$
is the reciprocal lattice constant.
\section{CONTINUITY AND INDEPENDENCE }
With the results of the previous section it is very easy to prove
that the spectra of $X$ and $P$ are continuous and that their
bases are unbiased. The continuity follows from the fact that
there is no restriction on the value that the parameters $a$ and
$g$ can take in $U_{a}$ and $V_{g}$ and therefore they can be
infinitesimal. So if $x$ is an eigenvalue for the position
operator then $x+\varepsilon$ with $\varepsilon\rightarrow 0$ is
also an eigenvalue (similarly for momentum eigenvalues). In order
to see that the basis $\{\varphi_{x}\}$ associated to $X$ and
$\{\phi_{p}\}$ associated to $P$ are unbiased we can first prove
that $|\langle\varphi_{x},\phi_{p}\rangle|=
|\langle\varphi_{x+s},\phi_{p}\rangle|$ for any $s$,  that is,
$|\langle\varphi_{x},\phi_{p}\rangle|$ is independent of $x$.
Indeed we have $ |\langle\varphi_{x},\phi_{p}\rangle| =
|e^{\frac{i}{\hbar}sp}\langle\varphi_{x},\phi_{p}\rangle|=
|\langle\varphi_{x},e^{\frac{i}{\hbar}sp}\phi_{p}\rangle|=
|\langle\varphi_{x},e^{\frac{i}{\hbar}sP}\phi_{p}\rangle|=
|\langle e^{\frac{-i}{\hbar}sP}\varphi_{x},\phi_{p}\rangle|=
|\langle\varphi_{x+s},\phi_{p}\rangle|$. In  the same way we see
that $ |\langle\varphi_{x},\phi_{p}\rangle|$ is independent of
$p$ and therefore the bases are mutually unbiased.

Actually, from the commutation relation we can get more than just
proving that $|\langle\varphi_{x},\phi_{p}\rangle|$ is constant;
we can get the dependence of the complex number
$\langle\varphi_{x},\phi_{p}\rangle$ on $x$ and $p$, that is, we
can calculate the coefficients of the unitary transformation
between the two bases. For this, let us consider
$\langle\varphi_{x+\varepsilon},\phi_{p}\rangle = \langle
U_{\varepsilon}\varphi_{x},\phi_{p}\rangle =
\langle\varphi_{x},U^{\dag}_{\varepsilon}\phi_{p}\rangle =
\langle\varphi_{x},\exp( \frac{i}{\hbar}\varepsilon
P)\phi_{p}\rangle = \exp( \frac{i}{\hbar}\varepsilon
p)\langle\varphi_{x},\phi_{p}\rangle$. With this we can build the
derivative
\begin{equation}\label{der}
  \frac{d\langle\varphi_{x},\phi_{p}\rangle}{dx}
  =\lim_{\varepsilon\rightarrow 0}
  \frac{\langle\varphi_{x+\varepsilon},\phi_{p}\rangle-
  \langle\varphi_{x},\phi_{p}\rangle}
  {\varepsilon} =\lim_{\varepsilon\rightarrow 0}
  \frac{\exp( \frac{i}{\hbar}\varepsilon
p)-1 }
  {\varepsilon} \langle\varphi_{x},\phi_{p}\rangle
  =\frac{i}{\hbar}p\langle\varphi_{x},\phi_{p}\rangle\ ,
\end{equation}
and integrating the equation we get
\begin{equation}\label{der2}
  \langle\varphi_{x},\phi_{p}\rangle = K \exp(\frac{i}{\hbar}xp)\
  ,
\end{equation}
with an arbitrary constant $K$.
\section{CONCLUSION}
We have seen that all essential properties of the position and
momentum observables as well as their relations arise in the
formalism of quantum mechanics from a unique feature: the
commutation relation $[X,P]=i\hbar$. However this fundamental
feature is a mathematical abstraction and it is therefore
convenient to emphasize its physical consequences. In the
teaching of quantum mechanics we can take two options: one can
postulate the commutation relations (postulate of canonical
quantization) and from this postulate to derive the physical
consequences, or, the reverse way, to define first the generators
of translations and boosts and derive the commutation relations
of the corresponding operators. The present paper is useful for
both strategies because it clearly presents the tight
relationship between the commutation relations and its physical
consequences. For this reason it was decided to present here a
complete proof, although some parts of it, but not all of them,
can be found spread in several texts.
\begin{acknowledgements}
This work received partial support from ``Consejo Nacional de
Investigaciones Cient\'{\i}ficas y T\'ecnicas'' (CONICET),
Argentina.
\end{acknowledgements}


\begin{thebibliography}{99}
\bibitem{boost} In this work we use the term ``boost'' with a
meaning related, but different from the one used in Galilei
transformations where the boosts are $X\rightarrow X+vt$ and
their generator is $B=Pt-mX$.
\end{thebibliography}
\end{document}